\documentclass[a4paper,10pt]{article} 
\usepackage[left=.65in,top=.65in,bottom=.65in,right=.65in]{geometry}
\usepackage{setspace} 
\usepackage[inline,shortlabels]{enumitem} 
\usepackage[super]{cite}

\usepackage{ragged2e}
\usepackage{amsmath}
\usepackage{amsthm}
\usepackage{amssymb}
\usepackage{caption}
\usepackage{orcidlink}
\usepackage{multicol}
\usepackage{nameref}
\usepackage{url}
\usepackage{lmodern}
\usepackage{graphicx} 
\usepackage{booktabs} 
\usepackage[T1]{fontenc}
\usepackage{helvet}         
\usepackage{setspace}

\usepackage{xcolor}

\usepackage[utf8]{inputenc} 
\usepackage{newunicodechar}
\newunicodechar{−}{-} 
\newunicodechar{́}{\'{}}
\usepackage{hyperref} 

\newcommand{\teff}{$T_{\mathrm{eff}}$}

\newcommand{\gaia}[0]{\emph{Gaia}}

\newcommand{\ha}{\mbox{H$\alpha$}}
\newcommand{\logg}{\mbox{$\log g$}}
\newcommand{\feh}{\mbox{\rm [Fe/H]}}
\newcommand{\msun}{\mbox{$\rm M_{\odot}$}}

\newcommand{\logp}{\mbox{$\textrm{log} P(\textrm{d})$}}
\renewcommand{\logp}{\mbox{$\log P$}}
\newcommand{\prot}{\mbox{$\textit{P}_{\textrm{rot}}$}}
\newcommand{\porb}{\mbox{$\textit{P}_{\textrm{orb}}$}}
\newcommand{\logrhk}{\mbox{$\log R^{\prime}_{\textrm{HK}}$}}
\newcommand{\primerhk}{\mbox{$R^{\prime}_{\textrm{HK}}$}}
\newcommand{\rhk}{\mbox{$R_{\textrm{HK}}$}}
\newcommand{\rhkphot}{\mbox{$R_{\textrm{HK,phot}}$}}
\newcommand{\shk}{\mbox{$S_{\textrm{HK}}$}}
\newcommand{\shklamost}{\mbox{$S_{\textrm{HK, LAMOST}}$}}
\newcommand{\shkmw}{\mbox{$S_{\text{HK, MW}}$}}

\newcommand{\aj}{Astron. J.}   
\newcommand{\apj}{Astrophys. J.}   
\newcommand{\apjl}{Astrophys. J. Lett.}   
\newcommand{\apjs}{Astrophys. J. Suppl. Ser.}   
\newcommand{\aap}{Astron. Astrophys.}   
\newcommand{\lrca}{Living Rev. Comput. Astrophys.}    
\newcommand{\lrsp}{Living Rev. Sol. Phys.}    
\newcommand{\mnras}{Mon. Not. R. Astron. Soc.}   
\newcommand{\ncomms}{Nat. Commun.} 
\newcommand{\pnas}{Proc. Natl Acad. Sci. USA}   
\newcommand{\raa}{Res. Astron. Astrophys.} 
\newcommand{\sci}{Science} 
\newcommand{\ssr}{Space Sci. Rev.}   

\makeatletter
\renewcommand{\maketitle}{\bgroup\setlength{\parindent}{0pt}

    \begin{Large} \begin{spacing}{1} \@title\end{spacing}\end{Large}
    \bigskip\linespread{1.2}
    \begin{normalsize} \@author \end{normalsize}

\bigskip
\@date
}
\makeatother

\newcommand{\beginsupplement}{%
    \setcounter{figure}{0}
}
\usepackage[symbol]{footmisc}

\title{\textbf{Enhanced magnetic activity in rapidly rotating binary stars}}
\author{%
Jie Yu\orcidlink{0000-0002-0007-6211}$^{1,2,3,4}$, 
Charlotte Gehan\orcidlink{0000-0002-0833-7084}$^{4}$, 
Saskia Hekker\orcidlink{0000-0002-1463-726X}$^{5,3}$, 
Mich\"ael Bazot\orcidlink{0000-0003-0166-1540}$^{3}$, 
Robert~H.~Cameron\orcidlink{0000-0001-9474-8447}$^{4}$, 
Patrick Gaulme\orcidlink{0000-0001-8330-5464}$^{6}$, \\
Timothy~R.~Bedding\orcidlink{0000-0001-5222-4661}$^{7}$, 
Simon~J.~Murphy\orcidlink{0000-0002-5648-3107}$^{8}$, 
Zhanwen Han\orcidlink{0000-0001-9204-7778}$^{9,10}$, 
Yuan-Sen Ting\orcidlink{0000-0001-5082-9536}$^{11,12}$, 
Jamie Tayar\orcidlink{0000-0002-4818-7885}$^{13,14}$, \\
Yajie Chen\orcidlink{0000-0001-5494-4339}$^{4}$, 
Laurent Gizon\orcidlink{0000-0001-7696-8665}$^{4,15}$, 
Jason Nordhaus\orcidlink{0000-0002-3138-8250}$^{16,17}$, 
Shaolan Bi\orcidlink{0000-0002-7642-7583}$^{18}$
}

\date{\footnotesize  
$^{1}$ School of Computing, Australian National University, Acton, ACT 2601, Australia.\\
$^{2}$ Research School of Astronomy \& Astrophysics, Australian National University, Cotter Rd., Weston, ACT 2611, Australia.\\
$^{3}$ Heidelberg Institute for Theoretical Studies (HITS) gGmbH, Schloss-Wolfsbrunnenweg 35, 69118 Heidelberg, Germany.\\
$^{4}$ Max-Planck-Institut f{\"u}r Sonnensystemforschung, Justus-von-Liebig-Weg 3, 37077 G{\"o}ttingen, Germany.\\
$^{5}$ Centre for Astronomy (ZAH/LSW), Heidelberg University, K{\"o}nigstuhl 12, 69117 Heidelberg, Germany.\\
$^{6}$ Th{\"u}ringer Landessternwarte, Sternwarte 5, 07778 Tautenburg, Germany.\\
$^{7}$ Sydney Institute for Astronomy, School of Physics, University of Sydney NSW 2006, Australia.\\
$^{8}$ Centre for Astrophysics, University of Southern Queensland, Toowoomba, QLD 4350, Australia.\\
$^{9}$ Yunnan Observatories, Chinese Academy of Sciences, Kunming 650216, China.\\
$^{10}$ International Centre of Supernovae, Yunnan Key Laboratory, Kunming 650216, China.\\
$^{11}$ Department of Astronomy, The Ohio State University, Columbus, OH 43210, USA.\\
$^{12}$ Center for Cosmology and AstroParticle Physics (CCAPP), The Ohio State University, Columbus, OH 43210, USA.\\
$^{13}$ Department of Astronomy, University of Florida, Bryant Space Science Center, Stadium Road, Gainesville, FL 32611, USA.\\
$^{14}$ Institute for Astronomy, University of Hawai`i, 2680 Woodlawn Drive, Honolulu, HI 96822, USA.\\
$^{15}$ Institut f{\"u}r Astrophysik, Georg-August-Universit{\"a}t G{\"o}ttingen, Friedrich-Hund-Platz 1, 37077 G{\"o}ttingen, Germany.\\
$^{16}$ Center for Computational Relativity and Gravitation, Rochester Institute of Technology, Rochester, NY 14623 USA.\\
$^{17}$ National Technical Institute for the Deaf, Rochester Institute of Technology, Rochester, NY 14623 USA.\\
$^{18}$ School of Physics and Astronomy, Beijing Normal University, Beijing 100875, People's Republic of China.
}

\begin{document}
\maketitle
\thispagestyle{empty}

\begingroup\footnotesize
\vspace{.3cm}
\endgroup


\vspace{3em}
\small
\justifying

\noindent\textbf{
Stellar activity is fundamental to stellar evolution and the formation and habitability of exoplanets. The interaction between convective motions and rotation in cool stars results in a dynamo process that drives magnetic surface activity. In single stars, activity increases with rotation rate until it saturates for stars with rotation periods $\prot < $ 3--10 d. However, the mechanism responsible for saturation remains unclear. Observations indicate that red giants in binary systems that are in spin-orbit resonance exhibit stronger chromospheric activity than single stars with similar rotation rates, suggesting that tidal flows can influence surface activity. Here, we investigate the chromospheric activity of main-sequence binary stars to understand the impact of tidal forces on saturation phenomena. For binaries with $0.5< \prot/\text{d} < 1$, mainly contact binaries that share a common thermal envelope, we find enhanced activity rather than saturation. This result supports theoretical predictions that a large-scale $\alpha-\omega$ dynamo during common-envelope evolution can generate strong magnetic fields. We also observe supersaturation in chromospheric activity, a phenomenon tentatively noted previously in coronal activity, where activity levels fall below saturation and decrease with shorter rotation periods. Our findings emphasise the importance of studying stellar activity in stars with extreme properties compared to the Sun's.
}

\vspace{3em}

Magnetic surface activity is prevalent in cool stars with convective outer layers. It manifests itself as various features in the chromosphere, such as bright plages surrounding starspots, dark filaments traversing the stellar disk, and prominences extending above the stellar limb\cite{basri2021}. Chromospheric magnetic activity can be measured from spectra: the emergence of the magnetic field, generated through a dynamo process in the stellar interior\cite{brun2017, cameron2023}, leads to chromospheric heating; this heating increases the emission in the cores of calcium absorption lines. Indeed, the so-called \shk\ index of Ca~II~H~\&~K emission is a well-known proxy of the chromospheric activity of the Sun\cite{egeland2017, sowmya2021} and other stars\cite{luhn2020, gehan2022}. Here, we use spectra from the \href{http://www.lamost.org/dr9/}{Ninth Data Release} (DR9) of the LAMOST survey\cite{cui2012} to measure the \shk\ index for a sample of \mbox{main-sequence} stars in binary systems (see Methods).

We select a sample of binaries from the Third Data Release (DR3) of \gaia, which provides 443,205 orbital solutions - including orbital periods, eccentricities, binary types, and filling factors used in this work — for 437,275 unique binary systems\cite{arenou2023}. To focus on main-sequence stars, we select 47,009 systems whose primary (brightest) star has \logg\ $>$ 3.5 dex. Here, we obtain \logg\ values from the \href{http://www.lamost.org/dr9/v2.0/}{LAMOST DR9v2.0 LRS Stellar Parameter Catalog} of AFGK stars, which is derived using the LAMOST stellar parameter pipeline (see Methods and Extended Data Fig.~\ref{fig:loggcomp} for details on the validation of the LAMOST \logg\ values for binary stars). We then measure the \shk\ index using the LAMOST spectra and calibrate it to the Mount Wilson \shk\ scale (see Methods and Extended Data Fig.~\ref{fig:sindexcalibratio}). To investigate the impact of tides on stellar magnetic activity, we exclude binaries with orbital periods (\porb) greater than 30 days, referred to as wide binaries in this study. These wide binaries are depicted by the lack of synchronisation (Extended Data Fig.~\ref{fig:rotation}). In these binary stars the \shk\ index is independent of the orbital period (Extended Data Fig.~\ref{fig:sp}), aligning with the absence of synchronisation. Therefore, these wide binaries can be regarded as single stars as far as their activity is concerned. Additional details on the activity of wide binaries and tidal synchronisation are provided in the Methods section. After applying these selection criteria, the retained binary systems are predominantly eclipsing and/or spectroscopic binaries. The primary stars with available \shk\ measurements span spectral types from K to early F and have eccentricities between 0 and 0.4.

\begin{figure}
\centering
\includegraphics[width=0.7\textwidth,clip]{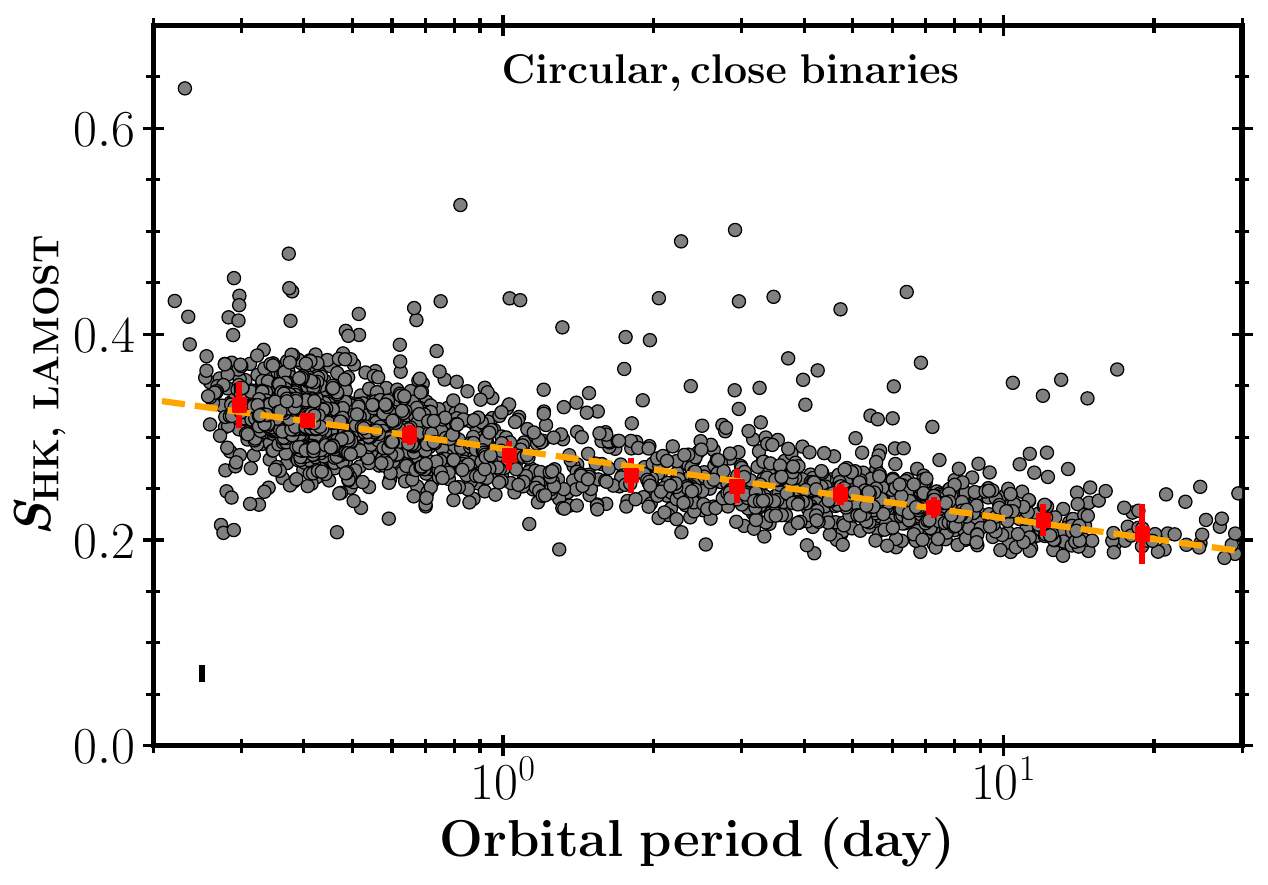}
\caption{\textbf{Chromospheric activity of close binaries on circular orbits.} Chromospheric activity, measured with the Ca II H \& K lines of LAMOST spectra and represented by the \shk\ index on the LAMOST scale (\shklamost), is shown as a function of the orbital period of 2079 \gaia\ binary stars on circular orbits (eccentricities less than 0.03). The red symbols depict the median \shk\ indexes of orbital period bins of $\delta \logp=0.2$, while the error bars indicate five times the standard errors. A linear fit to the median values is illustrated with a orange dashed line. The error bar in the lower left corner indicates the median uncertainty of the \shk\ index measurements for the entire sample, measured at $\sim$0.01.\label{fig:sp_cir}}
\end{figure}

Focusing on binaries with $\porb<30$ days, classified as close binaries in this study,  we find that at fixed orbital periods, systems on circular orbits (eccentricities $e < 0.03$) exhibit higher activity levels than those on eccentric orbits. This distinction is clearly demonstrated when comparing Fig.~\ref{fig:sp_cir} with Extended Data Fig.~\ref{fig:sp}, where the \shk\ indices of many eccentric systems approach the lower limit of 0.2. This suggests that circular binary stars are more likely to be synchronised (i.e., $\prot \approx \porb$). In fact, previous studies\cite{zahn1989a, barker2022, bashi2023} suggest that circularization generally occurs during the pre-main sequence phase, with a cutoff orbital period (\(P_{\text{cut}}\), where binaries with periods shorter than \(P_{\text{cut}}\) have circular orbits) between 7.2 and 8.5 days for stars with masses between 0.5 and 1.5~\msun\cite{zahn1989}. Given that these eccentric binaries with reduced activity are on the main sequence or subgiant branch (as identified by their \teff\ and \logg), and many have orbital periods below the cutoff period, it is likely that they have low-mass companions that cannot generate enough strong tidal forces to synchronize their primary stars\cite{lurie2017}. A recent study shows that $\sim$25\% of main-sequence binaries with orbital periods less than 6 days are on eccentric orbits\cite{dewberry2025}. However, the observation that synchronized stars tend to exhibit higher activity levels aligns with previous findings that red giants in spin-orbit resonance exhibit higher activity levels than those stars with similar rotation rates\cite{gaulme2014, gaulme2020, benbakoura2021, gehan2022, gehan2024}, known as RS CVn binaries\cite{hall1976}. Hence, our subsequent analysis focuses on binary systems with circular orbits unless otherwise stated.

As shown in Fig.~\ref{fig:sp_cir}, the \shk\ index for the close binaries increases with decreasing orbital period. Since the orbital period approximates the rotation period in these close binaries, this increase in the \shk\ index reflects the increase in activity observed in single stars, where more rapidly rotating stars exhibit higher activity until saturation is reached. Notably, the \shk\ index does not exhibit the expected saturation. This may be related to the inverse correlation between \teff\ and \shk, as well as the correlation between \teff\ and orbital period \cite{jayasinghe2020} (e.g., shorter orbital periods tend to be associated with smaller stars, which typically have lower \teff).

To eliminate the effect of \teff, we use another activity index, \logrhk, which measures the ratio of chromospheric flux associated with stellar activity to bolometric flux (see Methods for its definition). This metric mitigates colour-dependent effects on the \shk\ index and subtracts the photospheric flux contribution, allowing for a more accurate comparison of activity levels across different spectral types. Furthermore, to investigate the activity differences between single stars and binaries, we also examine the magnetic activity of 19,050 \textit{Kepler} and K2 single stars, with known \gaia\ and \textit{Kepler} binaries removed\cite{arenou2023, santos2019, santos2021}. The rotation periods of these stars have been precisely measured using high-precision light curves from the \textit{Kepler} and K2 space missions\cite{santos2019, santos2021, reinhold2020}, and their spectra are available in LAMOST DR9. This sample is supplemented with a group of rapid rotators, mainly from young open clusters\cite{wright2011,fang2018}, also observed by LAMOST.

\begin{figure}
\centering
\includegraphics[width=0.7\columnwidth]{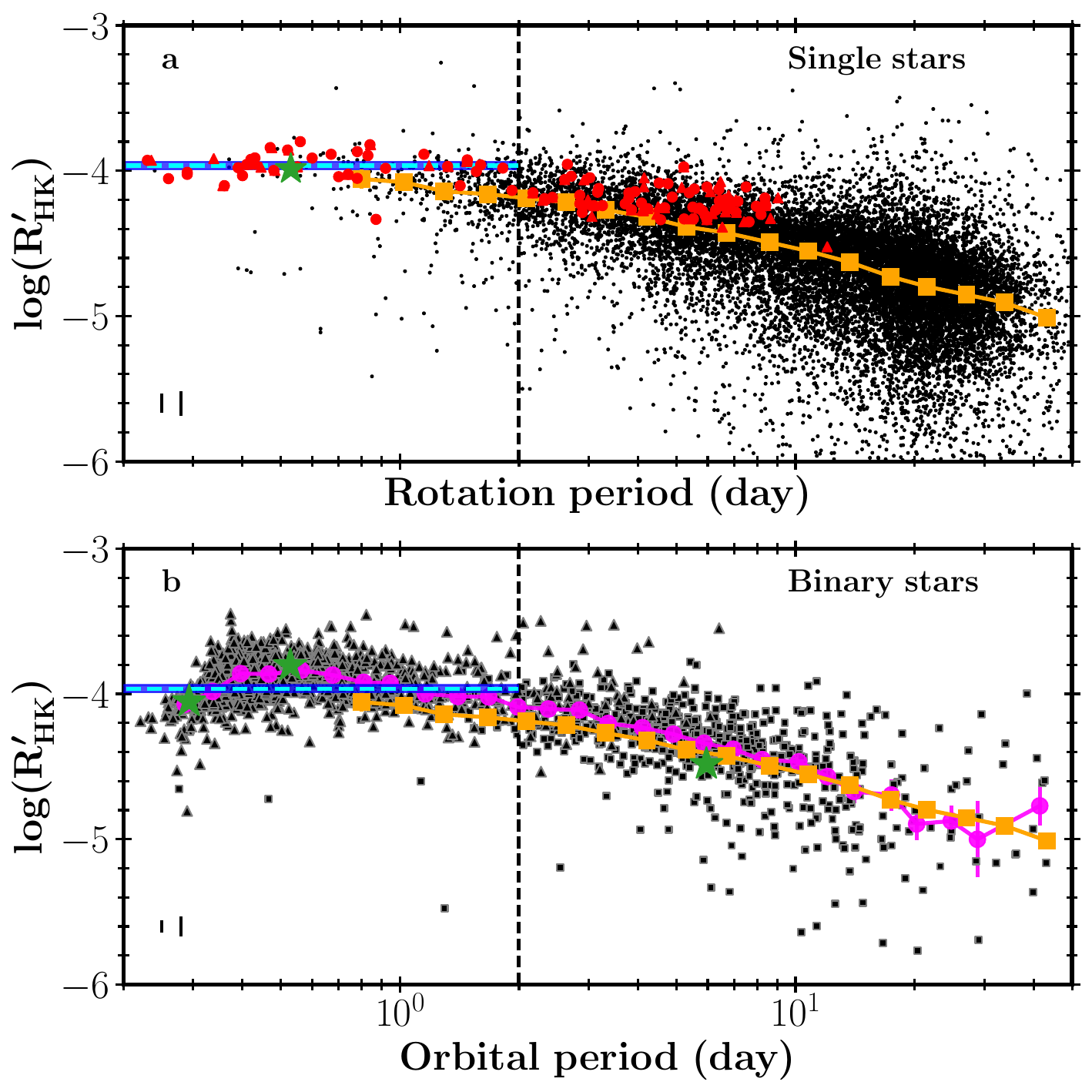}
\caption{\textbf{Comparison of magnetic activity between single and binary stars}. \textbf{Top}: Period–activity diagram for 16,844 \textit{Kepler}/K2 single stars. The vertical axis represents a chromospheric activity indicator, \logrhk, while the horizontal axis depicts the rotation periods whose values were adopted from the literature\cite{santos2019, santos2021, reinhold2020}. The red symbols represent a sample of reference stars observed by LAMOST and used to investigate the saturated and unsaturated activity with X-ray data by Wright et al. (2011)\cite{wright2011} (red squares) and H$_{\alpha}$ emission by Fang et al. (2018)\cite{fang2018} (red triangles). The vertical dashed line at $P = 2$ days indicates a conservative boundary below which magnetic activity has been widely established to saturate\cite{basri2021}. The mean \logrhk\ within the saturated regime for the reference stars is shown by the horizontal blue dashed line. The blue shaded region indicates twice the standard error of \logrhk. The orange squares connected by lines depict mean values within individual \logp\ bins of 0.11 width. The median formal \logrhk\ uncertainty and its conservative counterpart are shown in the lower-left corner (see Methods for more details). \textbf{Bottom}: Similar to the top panel, except for binary stars as a function of orbital period. We include 2095 \gaia\ circularised binary systems (eccentricities $<0.03$), among which 753 are spectroscopic binaries (squares) and 1342 are eclipsing binaries (triangles). The horizontal axis denotes the orbital period, and the magenta circles connected by lines depict mean values within \logp\ bins of 0.11 width, whose standard errors are too small to be visually discerned. The orange squares connected by lines, and dashed line are the same as in the top panel. The median formal and conservative \logrhk\ uncertainties for this binary star sample are also shown in the lower-left corner. The four single and binary stars marked with green asterisks represent stars exhibiting activity in the supersaturated, saturated, and unsaturated regimes, with their LAMOST spectra shown in Extended Data Fig.~\ref{fig:examplespectra}.\label{fig:singlevsbinary}}
\end{figure}

In Fig.~\ref{fig:singlevsbinary}a, we confirm that the activity levels of single stars increase as the rotation periods decrease, saturating the most rapidly rotating stars at $\logrhk \approx -4.0$. This overall trend is qualitatively consistent with previous studies across various activity indicators that probe stellar activity from the chromosphere to the corona\cite{noyes1984, wright2011, frasca2015, newton2017}. Furthermore, the saturated activity level determined in our analysis aligns precisely with the results of the literature using the same activity index based on the Ca II H \& K lines\cite{suarezmascareno2018, fang2018}. A subset of stars, shown in red symbols in Fig.~\ref{fig:singlevsbinary}a, exhibit higher overall activity than other stars, which is consistent with their young ages~($< 1$~Gyr\cite{wright2011, fang2018}). As these young stars age, they experience spin-down due to magnetic braking, resulting in decreased activity levels, particularly noticeable in the unsaturated regime.

In Fig.~\ref{fig:singlevsbinary}b, we show that stars in circularised binary systems with orbital periods greater than about 1 day exhibit activity levels comparable to those of single stars with the same rotation period\cite{nunez2024}. 
This similarity in activity levels between single stars and circularised binaries suggests that tidal interactions in these systems (i.e., orbital periods $\gtrsim$ 1 day) have a minimal direct effect on magnetic activity, aside from their role in synchronizing rotation periods.

However, as orbital periods decrease below $\sim$1 day, the magnetic activity of circularised binaries no longer saturates but instead continues to increase before declining for periods \( P \lesssim 0.5 \) days. To visualise this, Extended Data Fig.~\ref{fig:examplespectra}a presents two example LAMOST spectra near the Ca II H \& K lines, comparing a single star and a binary star with similar orbital and rotation periods, as well as comparable \teff\ and \logg. As expected, a more active star exhibits shallower absorption lines in its LAMOST lower-resolution spectrum, owing to activity-related emission, compared to a less active star. This enhanced activity highlights a complex interplay between rotation, tidal forces, and magnetic activity that differs from the well-studied saturated activity in single stars in this period regime. Although the well-known rotation-activity relation for single stars is proposed to be linked to a magnetohydrodynamic dynamo\cite{skumanich1972, noyes1984, pizzolato2003, wright2011, reiners2014}, the mechanism responsible for saturation remains unclear. A previous study\cite{reiners2014} suggests that the saturation could be caused by 1) a limit in the total energy that can be converted into stellar activity; 2) the entire surface being covered in spots; or 3) the quenching of differential rotation in fast rotators. Recently, Reiners et al. (2022)\cite{reiners2022} measured average magnetic fields for about 300 M dwarfs through Zeeman-broadening effects and found that the limit in available kinetic energy, rather than the available stellar surface area, sets the saturation of activity.

\begin{figure}
\centering
\includegraphics[width=0.6\columnwidth, clip]{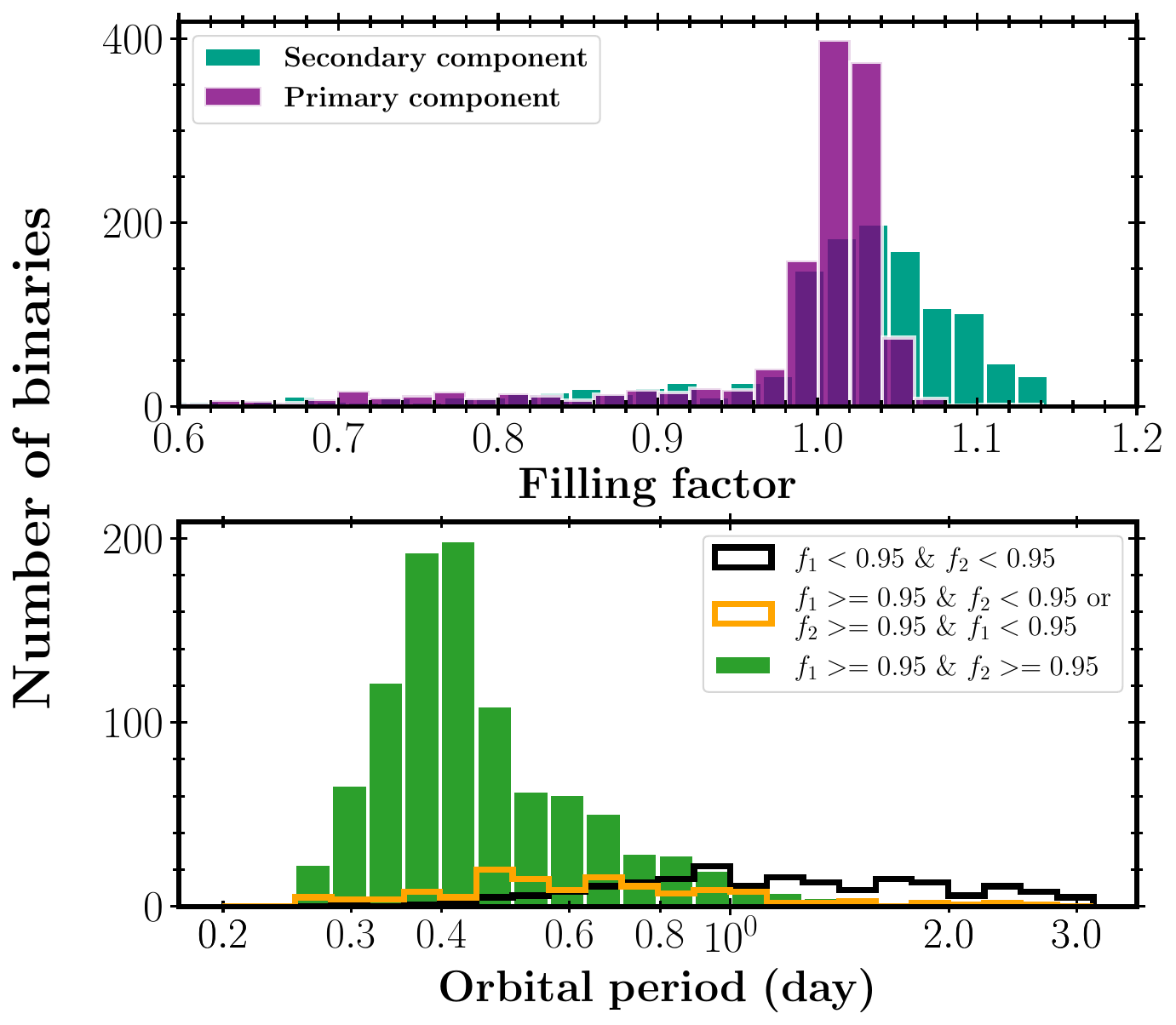}
\caption{\textbf{Filling factors for 1348 \gaia\ eclipsing binaries with \logrhk\ measurements}. \textbf{Top}: Histograms of filling factors, defined as the ratio of the stellar radius to the Roche lobe radius, for primary (\(f_1\), purple) and secondary (\(f_2\), cyan) components. \textbf{Bottom}: Orbital period distributions of detached (black), semidetached (yellow), and contact (green) binaries. The classification is based on filling factors: detached systems have \(f_1 < 0.95\) and \(f_2 < 0.95\), contact systems have \(f_1 > 0.95\) and \(f_2 > 0.95\), and semidetached systems fall into intermediate cases. The green histogram includes all contact systems with  \(f_1\), \(f_2\) $>$ 0.95  without a set upper limit. Among our sample, overcontact binaries, a subset of contact binaries, have filling factors reaching up to $\sim$1.14. Details on determining \gaia\ filling factors are provided in Arenou et al. (2023)\cite{arenou2023}. \label{fig:contactbinaries}}
\end{figure}

For binaries with $0.5 < \porb/\text{day} < 1.0$, the limit of kinetic energy can be altered due to mass transfer, which is accompanied by orbital energy and angular momentum transfer when binaries evolve from detached to semidetached and contact phases. Indeed, we find that 99\% of these circular binaries are eclipsing binaries, among which 7\%, 10\%, and 83\% are classified as detached, semi-detached, and contact binaries by \gaia, respectively (see Fig.~\ref{fig:contactbinaries}). Fig.~\ref{fig:contactbinaries} shows this classification based on the filling factors provided by \gaia\ \cite{arenou2023}, which are defined as the ratio of the stellar radius to the Roche lobe radius. In this work, we define detached systems as those with \(f_1 < 0.95\) and \(f_2 < 0.95\), contact systems as those with \(f_1 > 0.95\) and \(f_2 > 0.95\), and semidetached systems as intermediate cases where one component overfills or nearly fills its Roche lobe. Among our sample, overcontact binaries\textemdash a subset of contact binaries\textemdash exhibit filling factors of up to $\sim$1.14 (see Fig.~\ref{fig:contactbinaries}b). The dominance of contact binaries for \(\porb < 1\) days underscores the important role of tidal interactions and overflow in shaping the observed trends in activity.

It is noteworthy that this increase in the activity of binary stars ($0.5 < \porb/\text{day} < 1.0$) agrees with previous theoretical studies suggesting that a large-scale $\alpha-\omega$ dynamo during common envelope evolution (CEE) can generate strong magnetic fields\cite{regs1995, tout2008, potter2010, nordhaus2011}. The $\alpha$–$\omega$ dynamo converts kinetic energy into magnetic energy through two key effects: the $\omega$ effect, driven by differential rotation, transforms a poloidal magnetic field into a toroidal one, while the $\alpha$ effect, influenced by the Coriolis force, regenerates the poloidal field and sustains the magnetic field over time. Our observations support this theory, as the onset of enhanced activity at $\sim1$ day coincides with the orbital period threshold, where contact binaries become dominant over detached and semi-detached binaries. This finding provides valuable observational constraints on the role of magnetic fields during CEE, a phenomenon that remains poorly understood\cite{ropke2023}.

\begin{figure}
\centering
\includegraphics[width=0.7\columnwidth, clip]{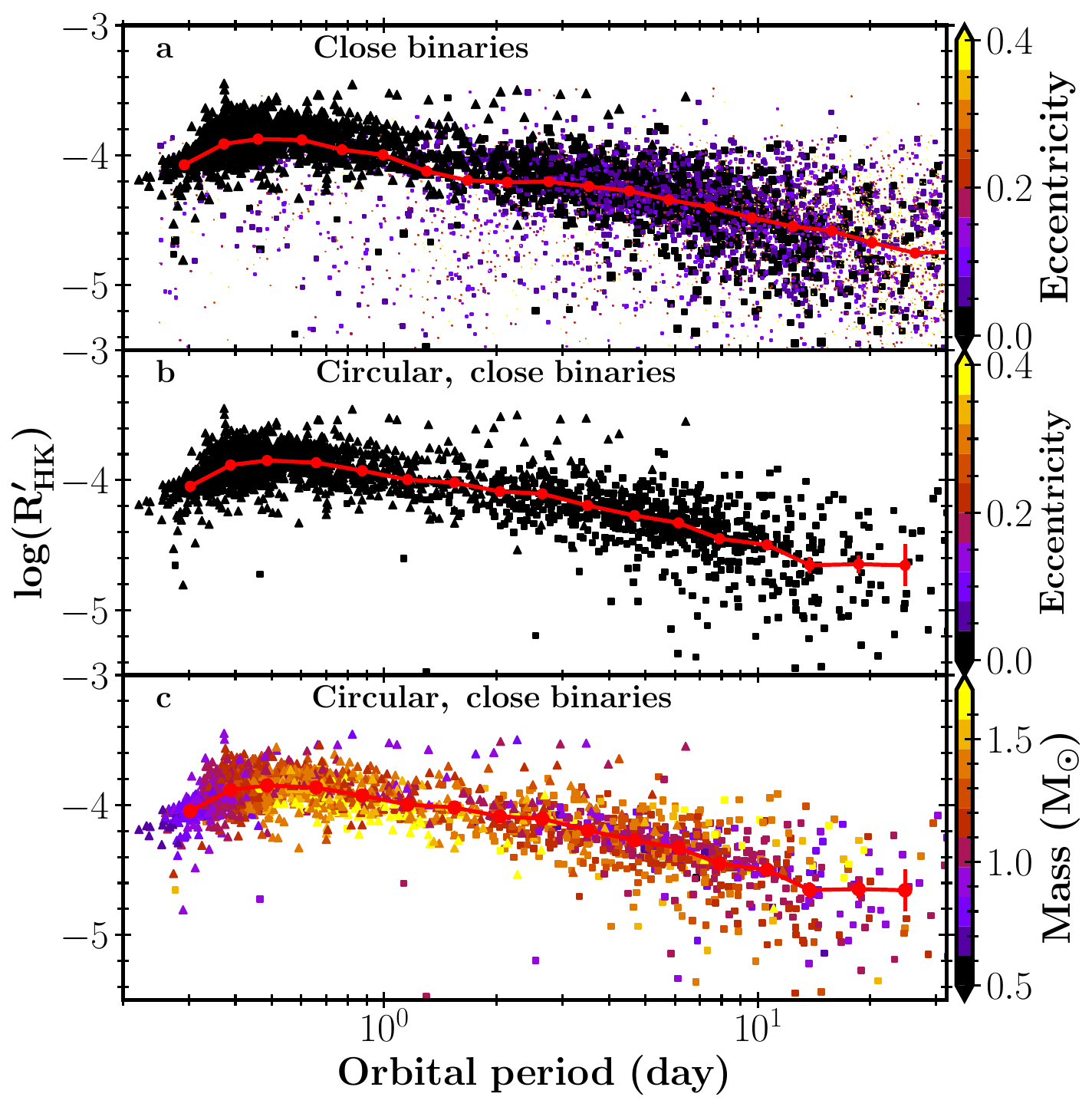}
\caption{\textbf{Chromospheric activity of close binaries and its relation with orbital eccentricity and stellar mass}. Panels \textbf{a} and \textbf{b} show \logrhk\ as a function of orbital period, color-coded by eccentricity, while panel \textbf{c} is color-coded by the mass of the primary. In all panels, triangles illustrate eclipsing binaries, and squares indicate spectroscopic binaries; red circles connected by lines represent mean values within \logp\ bins of 0.11 width; the standard errors of these means (shown with error bars) are mostly too small to be visually discerned. Panel \textbf{a} displays \gaia\ binaries with various eccentricities, where the size of each symbol is inversely proportional to eccentricity—smaller symbols correspond to more eccentric orbits. Panels \textbf{b} and \textbf{c} highlight those on circular orbits (eccentricity $<$ 0.03).\label{fig:eccentrictyMassyEffects}}
\end{figure}

Another intriguing observation is the decreased activity when \(\porb \lesssim 0.5\) days. This decline in chromospheric activity aligns with the previously identified supersaturation phenomenon, which has been tentatively observed in X-ray data as a measure of coronal activity in very rapidly rotating single stars\cite{randich1996, randich1998, wright2011}. To visualise supersaturation, Extended Data Fig.~\ref{fig:examplespectra}b presents three example LAMOST spectra near the Ca II H \& K lines, comparing three binary stars with similar \teff\ and \logg\ values across the unsaturated, saturated, and supersaturated regimes. To further investigate supersaturation, Fig.~\ref{fig:eccentrictyMassyEffects} illustrates the relationship between \logrhk, orbital period, eccentricity, and primary mass (see Methods for mass determination) for close binary systems. This figure reveals that supersaturated binaries are characterised by short orbital periods, low eccentricities, and low primary masses ($<$ 1.25\msun).

Supersaturation is even less well understood than the phenomenon of saturation. Supersaturation was first reported by Randich et al. (1996)\cite{randich1996} for rapidly rotating stars ($P \lesssim 0.5$ days) using X--ray data. Subsequent studies found tentative evidence for reduced magnetic activity in very rapidly rotating M dwarfs, observed through X-ray emissions\cite{wright2011, nunez2024} and flaring activity\cite{gunther2020, ramsay2020}. In Fig.~\ref{fig:singlevsbinary}b, we provide unambiguous evidence for supersaturation in chromospheric indices, thanks to a large sample of rapidly rotating binary stars. Our detection of supersaturation in the fastest rotators indicates that the whole outer layer of the atmosphere is affected. Identifying the mechanism behind this supersaturation remains difficult because these stars rotate about 50 times faster than the Sun, which forms the basis of our current understanding of stellar dynamos and the interaction between magnetic fields and stellar atmospheres. Our results highlight the need for improved observations of stars with properties that are extreme compared to the Sun’s --- understanding these extreme cases will shed light on the underlying physical mechanisms.

\section*{Methods}
\noindent\textbf{\shk\ Index Measurements} We employ a standard method, as outlined in previous studies\cite{zhang2020, gomesdasilva2021, gehan2022}, to calculate the \shk\ index, a widely used primary indicator of stellar chromospheric activity. We cross-match the binary systems from \gaia\ DR3 with those from the 9th Data Release (DR9) of the LAMOST survey, using the internal \gaia\ source IDs present in both catalogues. This yields 76,178 spectra available for 48,347 unique \gaia\ binary systems. We focus on spectra with SNRs in the u band greater than 5 (uSNRs provided by LAMOST) and remove M-type binary stars from our sample using the spectral-type classifications in the LAMOST DR9v2.0 database. This is required by our requirement for \teff\ values from this data release, which is optimised for A, F, G, and K stars. Additionally, we encounter higher noise levels in the Ca II H \& K lines of M-type stars, which are located at the blue end of the spectral energy distributions, complicating the precise measurement of the \shk\ index.

To measure the \shk\ index, we first determine the radial velocity of a spectrum to locate the Ca II H \& K lines, using a method similar to previous studies\cite{gehan2022, yu2024}. We search for the strongest line of either \ha\ ($\lambda_0=6563$ \AA\ in air) or Na D lines ($\lambda_0=5893$ \AA\ in air) within a 150 \AA\ window centered at their respective wavelengths among the LAMOST spectra. We then fit a composite model of a line and a Gaussian to the detected lines and use the central Gaussian wavelength $\lambda$ to calculate the radial velocity, $v = c*(\lambda-\lambda_0)/\lambda_0$, where $c$ is the speed of light in vacuum. Our final radial velocity estimate is taken from the strongest line only if the absorption depth of the line is at least 6\%.

After identifying the Ca II H \& K lines using the radial velocities determined above, we calculate the integrated emission line fluxes for the H and K bandpasses using a triangular window centred at the line cores ($\lambda_0 = 3968$~\AA~and 3934 \AA, respectively). Differing from the FWHM = 1.09~\AA\ window employed in previous studies for the analysis of high-resolution HARPS spectra\cite{gomesdasilva2021,marvin2023}, we opt for a broader window (4.36 \AA) to fully capture the entire activity emission present in the Ca II H \& K lines within lower-resolution LAMOST spectra. We note that our main conclusions remain unchanged, as tested with FWHM values of 1.09, 2.18, and 4.36 \AA. Another reason for choosing FWHM = 4.36~Å is that our final activity index measurements (\rhk, see the following section for its definition) for single stars in the saturated regime are in good agreement with the values of the literature at \rhk = -4.0 (see Fig.~\ref{fig:singlevsbinary}). 

We determine the continuum fluxes for the R and V bandpasses using 20 \AA\ rectangular windows centered at 4001 \AA\ and 3901 \AA, respectively. The \shk\ index is calculated as the flux ratio and modified according to Karoff et al. (2016)\cite{karoff2016}:

\begin{equation}
\shk\ = \alpha \times 8 \times \frac{4.36~\text{\r{A}}}{20~\text{\r{A}}} \times \frac{H+K}{R+V},
\end{equation}
where $\alpha=1.8$. To estimate the uncertainty of the \shk\ index, we also follow the method proposed by Karoff et al. (2016)\cite{karoff2016}, where $\log{\sigma(\shk)}=-\log(\overline{S/N})-0.5$. Here, $\overline{S/N}$ is the average SNR as mentioned earlier and can be obtained from LAMOST spectrum files. In cases where multiple spectra are available, we use the mean value and the mean uncertainty. We refer to these \shk\ index measurements as $S_{\textrm{HK, LAMOST}}$, which are on a LAMOST scale. Using this method, we measure $S_{\textrm{HK, LAMOST}}$ for 31,099 binary stars.\\

\noindent\textbf{\shk\ Index Calibration} We calibrate our \shklamost\ index measurements to the Mount Wilson scale (\shkmw) using 273 stars (\gaia\ $G<13$) in common from literature studies\cite{isaacson2010, borosaikia2018, gomesdasilva2021, isaacson2024}. The best linear fit, after iterative rejection of 11 3$\sigma$ outliers (1 iteration after convergence), is $\shkmw=(5.132\pm0.276) \times S_{\textrm{HK,~LAMOST}} -(0.865\pm0.066)$. These calibration stars span a wide range of spectral types, from K to F (see Extended Data Fig.~\ref{fig:sindexcalibratio}). The median formal uncertainty of the calibrated \shklamost\ index measurements for the circular, close binary stars is 0.03. The standard deviation between our calibrated \shklamost\ measurements and the literature values is 0.08, which includes formal and calibration uncertainties and the scatter due to activity variations because the LAMOST spectra were collected at different epochs compared to those used in the literature studies\cite{isaacson2010, borosaikia2018, gomesdasilva2021, isaacson2024}.
\bigskip

\noindent\textbf{Tidal synchronisation}
Extended Data Fig.~\ref{fig:rotation} shows that the rotation and orbital periods are not correlated for the binaries with $\porb \gtrsim 30$ days. This independency suggests that these binaries on wide orbits are not synchronised through tidal interactions. As such, for these wide binaries, the \shk\ index is independent of the orbital period (see Extended Data Fig.~\ref{fig:sp}).

Extended Data Fig.~\ref{fig:sp} also shows substantial dispersion in the \shk\ indexes of wide binary systems. This dispersion arises from various factors, including stellar characteristics (e.g., effective temperatures \teff), evolutionary stages, and rotation rates\cite{gomesdasilva2021}, as well as intrinsic variability due to short-term stellar rotation and longer-term magnetic cycles\cite{jeffers2023}. To compare with solar \shk, we convert our \shk\ index measurements to the Mt. Wilson scale (see Methods). The mean value and standard error of the calibrated \shk\ indexes for the solar analogs ($\Delta$\teff$<100$K, $\Delta \logg<0.1$, and $\Delta \feh<0.1$) in the wide binaries is $0.231\pm0.004$, with \teff, \logg, and \feh\ values from the \href{http://www.lamost.org/dr9/v2.0/doc/lr-data-production-description#S3.2}{LAMOST DR9}. This is higher than the solar value during solar cycle23, $<S_{23}>=0.1701\pm0.0005$\cite{egeland2017}, consistent with prior studies indicating that the Sun is less active than other Sun-like stars\cite{reinhold2020a, zhang2020a}.

We find that 87\% of the close binaries with $\porb \lesssim 30$ days are in synchronous rotation, that is, the orbital and rotation periods do not differ by more than a factor of 2. We note that while the rotation periods derived from Kepler/K2 data are constrained by the 90-day duration of typical one-quarter light curves, this limitation does not introduce biases for the orbital period threshold. Within this sample of synchronised systems, latitudinal differential rotation\cite{lurie2017} and starspot evolution\cite{basri2020} may account for some of the scatter for $\prot<\porb<2\prot$ (above the dashed line in Extended Data Fig.~\ref{fig:rotation}), and for systems with $0.5\prot<\porb<\prot$ (below the dashed line in Extended Data Fig.~\ref{fig:rotation}), the synchronization process might still be ongoing. There is an over-density and a pronounced reduction in scatter in shorter-period binaries ($P_{\textrm{orb}} < 10$ days), which suggests increased tidal forces. The 10.5\% of systems in this regime that are not synchronised can be attributed to their low-mass companions (\(M_2 \sin i < 0.06~\msun\), where \(i\) is the orbital inclination), which generate weak tidal forces insufficient to synchronise the primary stars\cite{lurie2017}. Here, \(M_2 \sin i\) measurements are available for six systems, with values inferred from the mass functions provided in the \gaia\ orbital solutions catalog, combined with the primary masses estimated in this work. Alternatively, this lack of synchronization could be explained by young ages or a complex dynamical history\cite{lurie2017}. \\

\noindent\textbf{\logrhk\ Measurements} To better compare the activity of stars across different spectral types\cite{gomesdasilva2021}, we calculate the ratio of the chromospheric fluxes in the Ca II H \& K lines with respect to the bolometric flux, known as \logrhk. Specifically, the parameter \primerhk\ is computed as follows:
\begin{equation}
\primerhk = \rhk - \rhkphot,
\end{equation}
where $R_{\textrm{HK}}$ represents the chromospheric surface flux ratio (with respect to the bolometric flux), and $R_{\textrm{HK,phot}}$ denotes the photospheric flux ratio. This subtraction minimises flux contamination from the photosphere, particularly for late F-type stars in our sample. The method pioneered by Noyes et al. (1984)\cite{noyes1984} calculates $R_{\textrm{HK}}$ as:
\begin{equation}
R_{\textrm{HK}} = 1.34 \times 10^{-4}~C_{\textrm{cf}}~\shk,
\end{equation}
where $C_{\textrm{cf}}$ is a color-dependent conversion factor to remove the color dependence of the \shk\ index. Following the scheme of Rutten et al. (1984)\cite{rutten1984}, we use:
\begin{equation}
\log C_{\textrm{cf}} = 0.25(B-V)^3 - 1.33(B-V)^2 + 0.43(B-V) + 0.24
\end{equation}
for dwarf stars with $0.3<(B-V)<1.6$. We then calculate \rhkphot\ as per Hartmann et al. (1984)\cite{hartmann1984} and Noyes et al. (1984)\cite{noyes1984}:
\begin{equation}
\log \rhkphot = -4.898 + 1.918 (B-V)^2 - 2.893(B-V)^3.
\end{equation}

Given that the \gaia\ binary stars are unresolved, instead of using $B$ and $V$ magnitudes from photometric surveys, we calculate $(B-V)$ from spectroscopic \teff\ provided by LAMOST DR9 to minimise contamination from companions (see the validation of LAMOST \teff\ values below). For this, we adopt $(B-V)$ and \teff\ values from Huang et al. (2015)\cite{huang2015} for 134 dwarf stars with $3100~\textrm{K}<$\teff$<9700~\textrm{K}$ and fit a quadratic polynomial, which is optimised as:
\begin{equation}
(B-V) = 3.964 * (T_{\textrm{eff}}/10^4)^2 - 7.709 * (T_{\textrm{eff}}/10^4) + 3.738.
\end{equation}
Finally, we take the logarithmic scale of \primerhk, namely \logrhk, to investigate the magnetic activity of stars with different spectral types.

We estimate two sets of \logrhk\ uncertainties through Monte Carlo simulations: formal and conservative. For the formal (statistical) uncertainty, we perturb \teff\ and \shk\ 1000 times assuming Gaussian errors. In each iteration, we recalculate \logrhk\ using the perturbed \teff\ and \shk\ values and take the standard deviation as the formal \logrhk\ uncertainty. For the conservative \logrhk\ uncertainty, we use a similar method but conservatively approximate the \shk\ uncertainty with the standard deviation of the difference between the calibrated and literature \shk\ indexes for 273 standard stars shown in Extended Data Fig.~\ref{fig:sindexcalibratio}. These conservative \logrhk\ uncertainties account for the activity variations on the time scales of stellar rotation and activity cycles, in addition to the statistical uncertainties and the \shk\ calibration uncertainties. The formal and conservative \logrhk\ uncertainties are 0.066 and 0.085, respectively, for the \textit{Kepler}/K2 rotator sample (shown in the lower left corner of Fig.~\ref{fig:singlevsbinary}a), and 0.041 and 0.068 respectively, for the binary star sample (shown in the lower left corner of Fig.~\ref{fig:singlevsbinary}b).

Finally, we note that although our \shklamost\ values, and consequently \logrhk\ values, are calibrated using brighter stars ($G<13$), this calibration are expected to be valid when applied to fainter stars in our sample. Notably, the observed trends in the unsaturated, saturated, and supersaturated regimes (see Fig.2) remain consistent when the sample is restricted to stars brighter than 13th magnitude in the \gaia\ G-band (see Supplementary Fig. 1). This consistency underscores the reliability of our analysis, as the trends are less impacted by noise in the LAMOST spectra.\\

\noindent\textbf{Binary Star Mass Measurements} We determine the masses of the \gaia\ binary stars using the \texttt{isoclassify} method\cite{huber2017,berger2020}. This approach undertakes grid-based modeling by employing the \texttt{MIST} stellar tracks\cite{dotter2016, choi2016}. Our observational constraints comprise spectroscopic values for \teff, \logg, and [Fe/H] adopted from the LAMOST DR9v2.0 LRS Stellar Parameter Catalog. These values are derived from low-resolution spectra obtained from the same source in the LAMOST database. While incorporating supplementary data of photometry and parallaxes can generally enhance the precision of stellar mass estimates for single stars, we refrain from doing so for \gaia\ binary stars due to their unresolved nature. Otherwise, the inclusion of photometry and parallax data in such cases can introduce biases into the mass estimates. Lastly, we only use mass measurements with a precision better than 30\% in our analysis.

\begin{figure}[ht]
\centering
\includegraphics[width=0.9\textwidth]{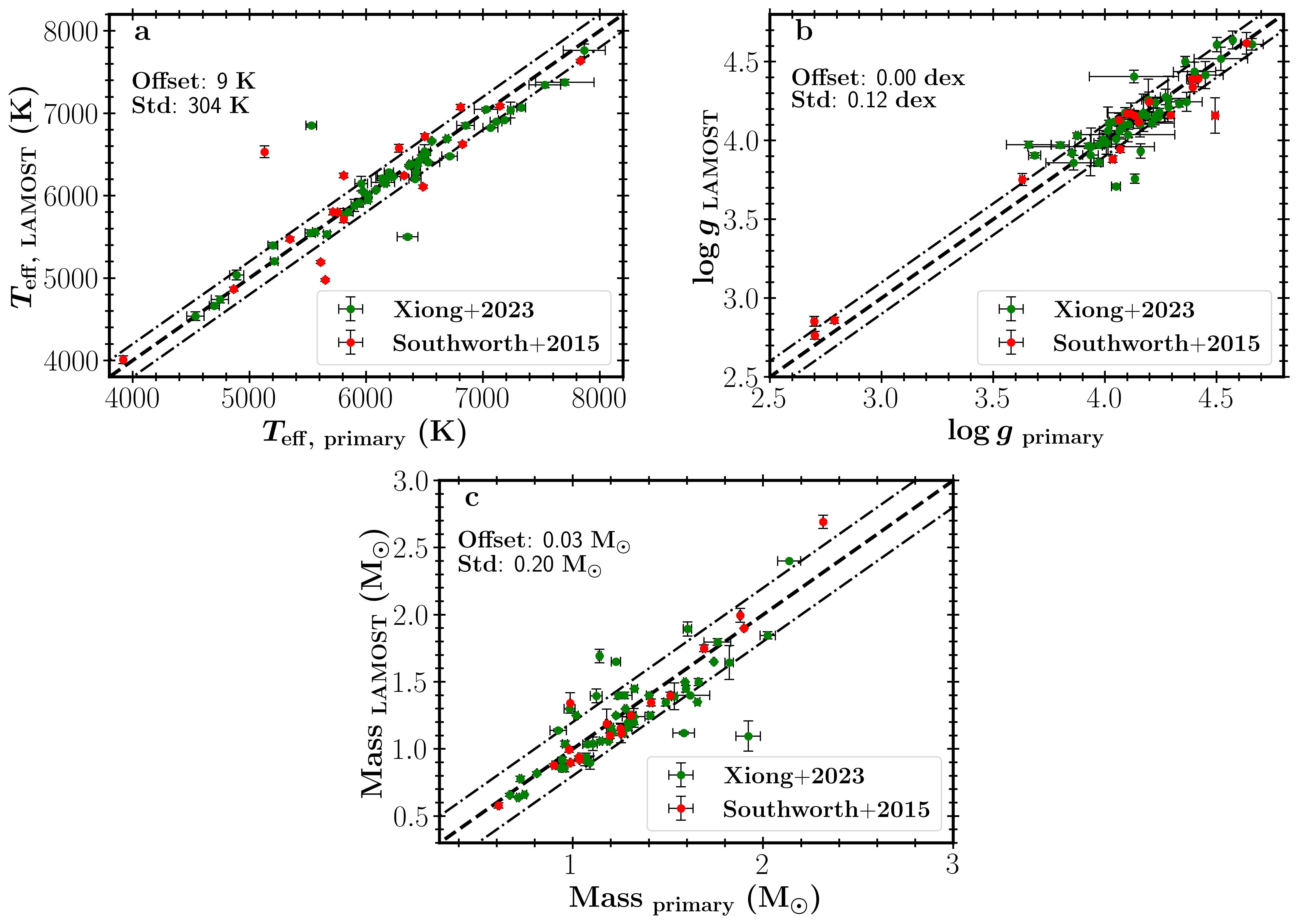}\\
\caption{\textbf{Validation of LAMOST stellar parameters using double-lined spectroscopic binaries}. Panels \textbf{a}, \textbf{b}, and \textbf{c} compare \teff, \logg, and stellar masses obtained from the LAMOST DR9v2.0 Stellar Parameters Catalog with literature values\cite{southworth2015,xiong2023} for the primary stars in SB2 systems. LAMOST \teff\ and \logg\ values (vertical axes) were obtained from single-star atmospheric models applied to LAMOST spectra. Stellar masses were determined by isochrone fitting using \texttt{isoclassify}, with LAMOST \teff, \logg, and [Fe/H] as input constraints (see Methods for details). Error bars in each panel represent the 1-$\sigma$ uncertainties. Along the three horizontal axes, the \teff\ values for the primary stars obtained from the literature studies are estimated from spectra observed near the secondary minimum eclipses to minimise flux contamination from companions. The mass values are dynamical masses of SB2 stars derived from radial velocities and light curves, while the \logg\ values are calculated from the dynamical masses and radii of SB2 binaries. These comparisons show an offset of 9 K with a standard deviation of 304 K for \teff, 0.00 dex with a standard deviation of 0.12 dex for \logg, and 0.03~\msun\ with a standard deviation of 0.20~\msun\ for mass. The black dashed line in each panel represents the one-to-one relation, while the dashed-dotted lines indicate offsets of 200~K for \teff, 0.12 dex for \logg, and 0.2~\msun\ for mass.\label{fig:loggcomp}}
\end{figure}

\begin{figure}
\centering
\includegraphics[width=0.7\columnwidth, clip]{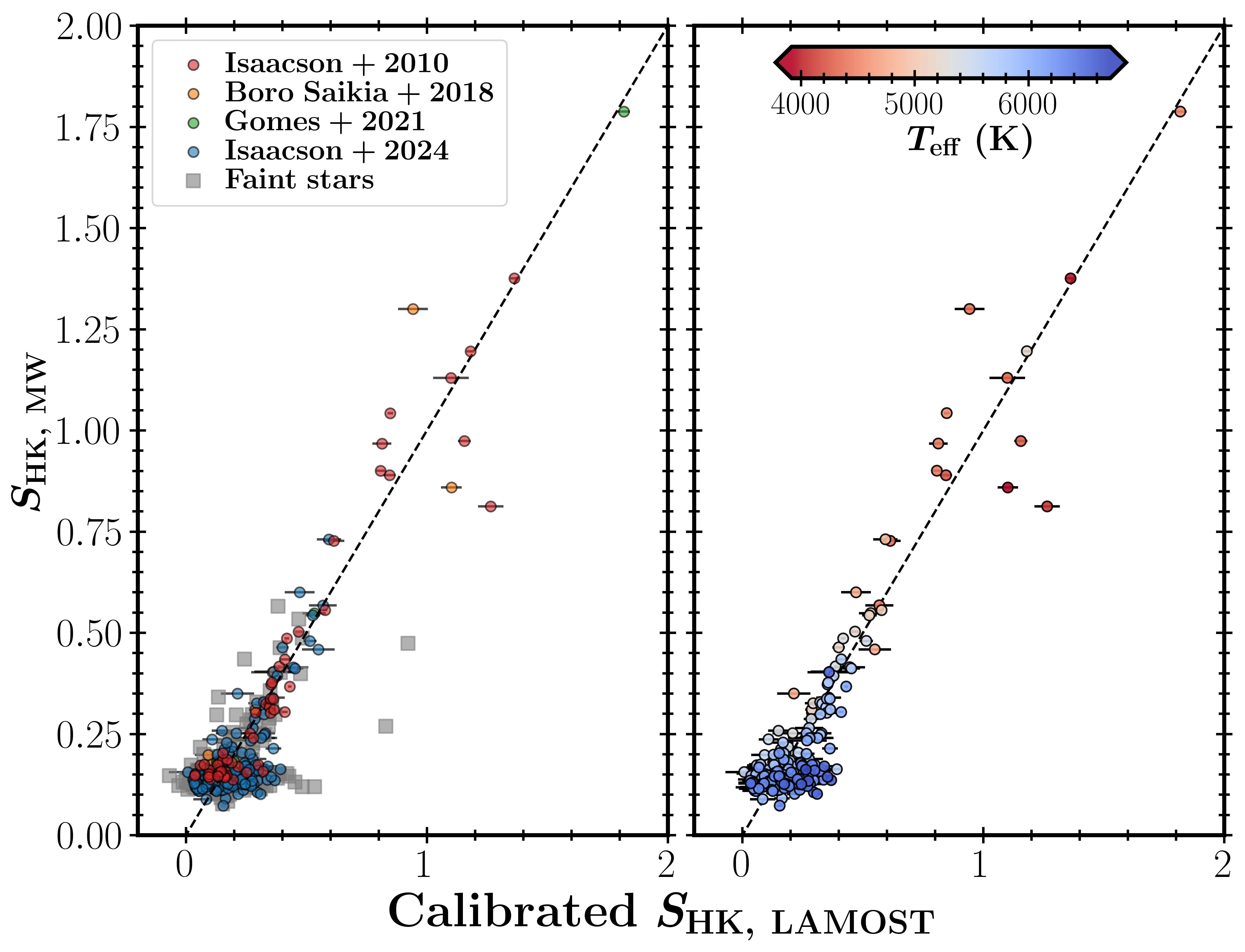}
\caption{\textbf{Calibration of the \shk\ index}. The calibration sample consists of 273 stars brighter than 13th magnitude in the \gaia\ \(G\)-band (circles in both panels), for which the \shk\ index values on the Mount Wilson scale ($\shk_{\text{,~MW}}$) are available from the literature\cite{isaacson2024, gomesdasilva2021, borosaikia2018, isaacson2010}. We performed an iterative linear fitting by rejecting 3-$\sigma$ outliers (11 stars removed, 1 iteration after convergence). The resulting best fit is presented with the black dashed line: $\shkmw=(5.132\pm0.276) \times S_{\textrm{HK,~LAMOST}} -(0.865\pm0.066)$, where $S_{\textrm{HK,~LAMOST}}$ represents the \shk\ index measurements from our work. The 1-$\sigma$ formal uncertainties of the calibrated \shk\ values are shown as error bars in both panels. The faint stars (\(G > 13\), squares shown in the left panel), which were not included in the calibration, contribute to the large scatter observed at low \(S_{\textrm{HK,~LAMOST}}\) values. This is expected, as the spectra of stars with small \shk\ values are more likely to be affected by noise. The left panel highlights the literature subsamples used for the calibration, while the right panel is color-coded by the \teff\ range of the calibration sample, with \teff\ values obtained from LAMOST DR9.\label{fig:sindexcalibratio}}
\end{figure}

\begin{figure}
\centering
\includegraphics[width=0.7\columnwidth]{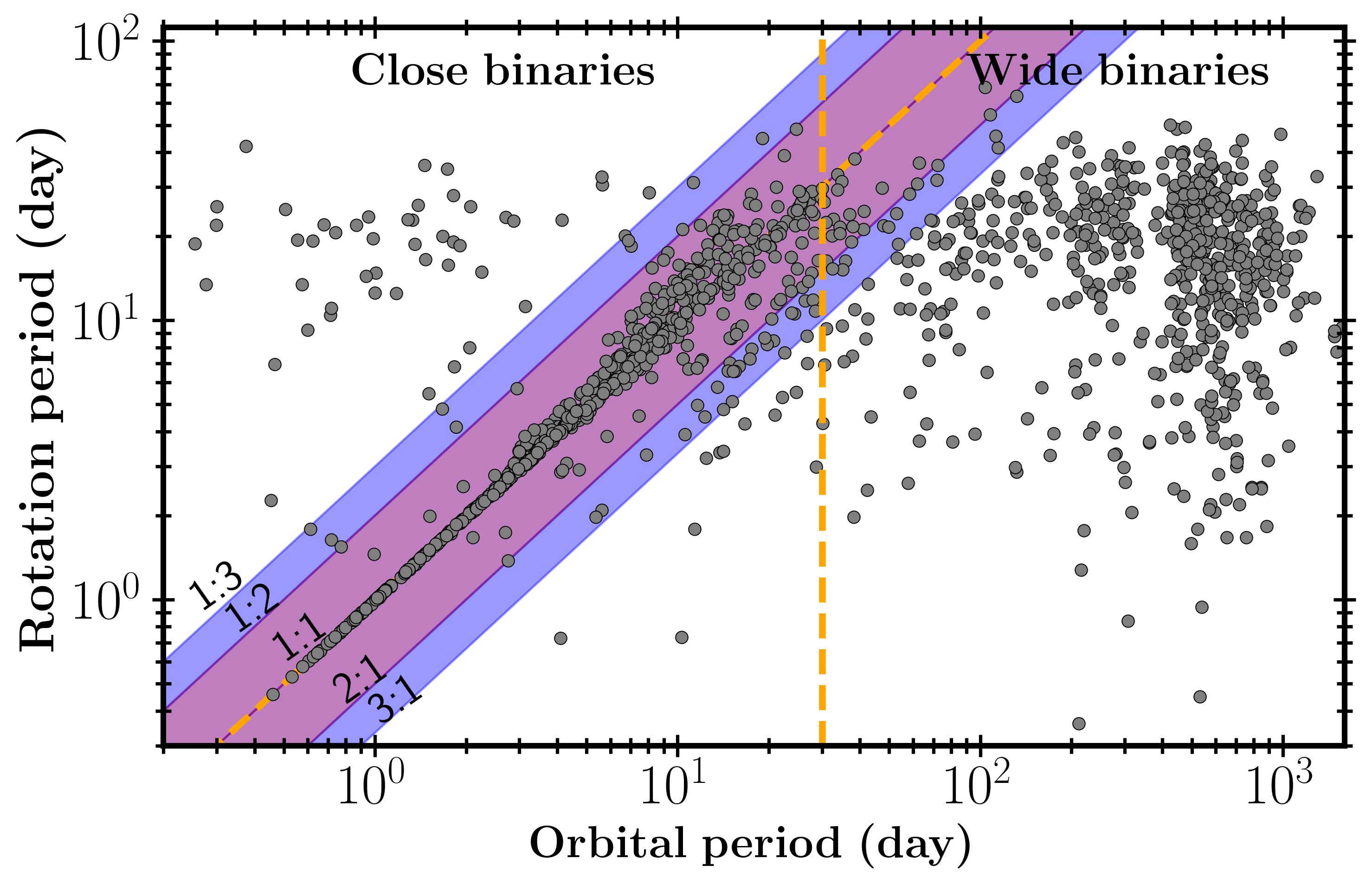}
\caption{\textbf{Comparison of rotation periods (\prot) and orbital periods (\porb) for 1,407 main-sequence and subgiant stars}. This sample includes \textit{Kepler} eclipsing binaries\cite{lurie2017} and \gaia\ binaries\cite{arenou2023}, whose rotation periods are available from previous studies\cite{santos2019, reinhold2020, santos2021} based on \textit{Kepler} and K2\cite{reinhold2020} light curves. The purple shading indicates the region where the \prot/\porb\ ratio falls between 1:2 and 2:1, while the blue shading highlights the region where the \prot/\porb\ ratio is between 2:1 and 3:1, or between 1:3 and 1:2. The diagonal dashed line represents a one-to-one correspondence. The vertical dashed line at $P=30$ days signifies a boundary beyond which binaries are almost not synchronised. We categorise the systems to the left as close binaries and those to the right as wide binaries.\label{fig:rotation}}
\end{figure}

\begin{figure}
\centering
\includegraphics[width=0.7\textwidth,clip]{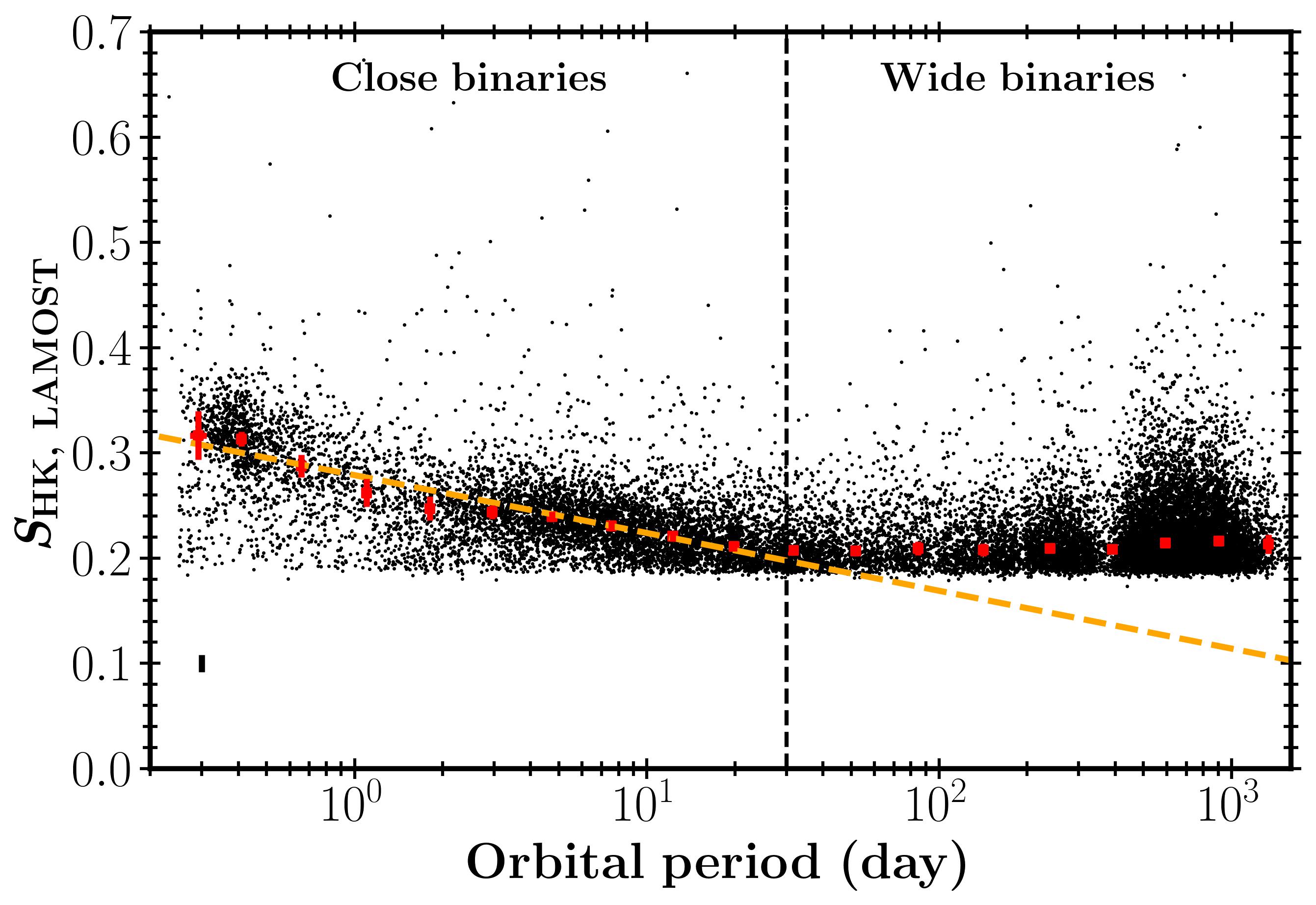}
\caption{\textbf{Chromospheric activity of close and wide \gaia\ binary systems}. The \shklamost\ values are measured from the Ca II H \& K lines in LAMOST spectra and expressed as the \shk\ index on the LAMOST scale. The vertical dashed line at $P=30$ days serves as a boundary distinguishing the regime where the \shk\ index depends on the orbital period ($\porb \lesssim 30$ days) from the regime where it does not ($\porb \gtrsim 30$ days). The red symbols depict the median \shk\ indexes of orbital period bins of $\delta \logp=0.2$, while the error bars indicate five times the standard errors. A linear fit to the median values of the close binaries is illustrated with an orange dashed line. The error bar in the lower left corner indicates the median uncertainty of the \shk\ index measurements for the entire sample, measured at $\sim$0.01.\label{fig:sp}}
\end{figure}

\begin{figure}
\centering
\includegraphics[width=0.7\textwidth,clip]{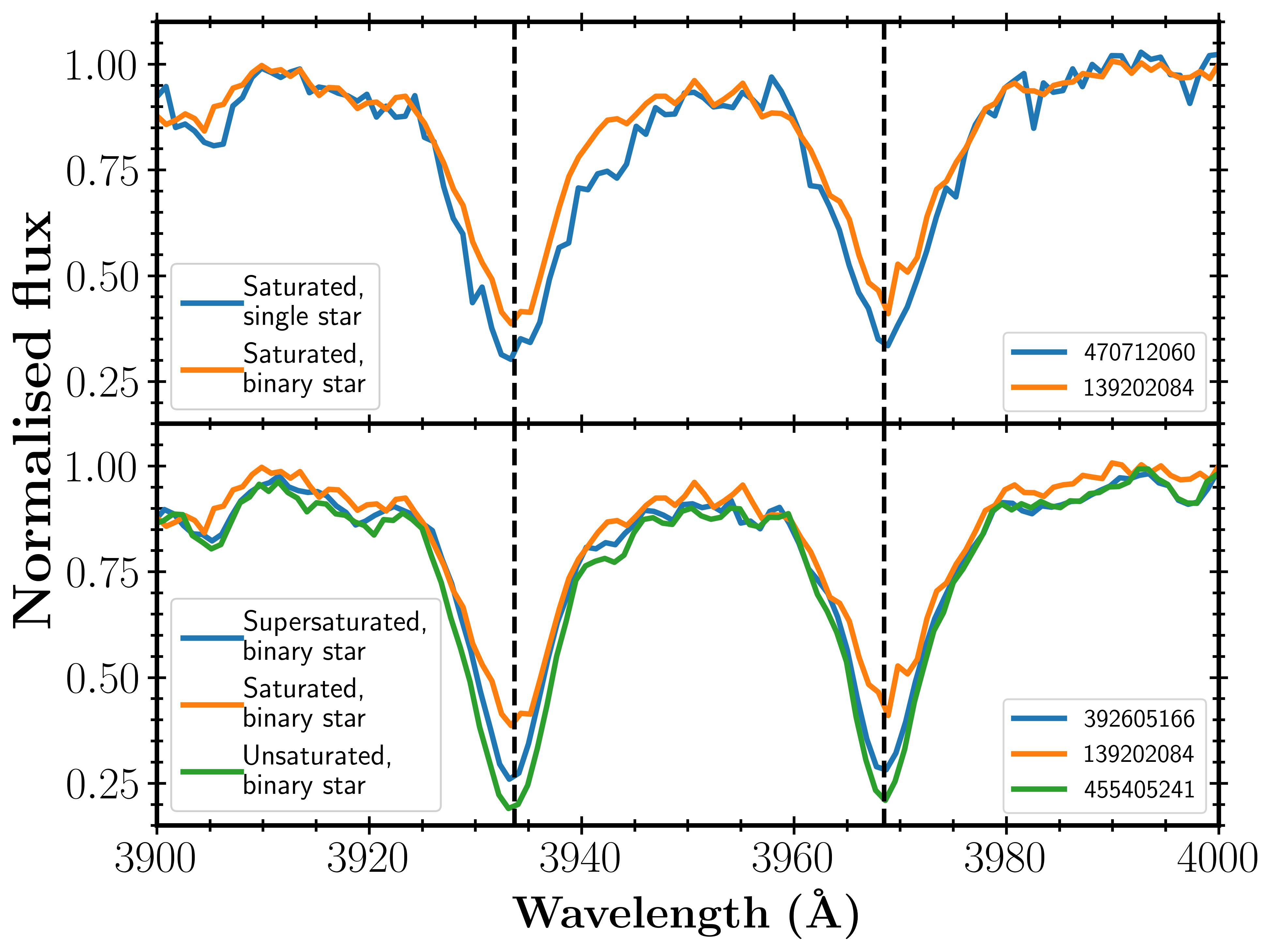}
\caption{\textbf{Five example LAMOST low-resolution spectra around the Ca II H \& K lines}. The wavelengths are shifted to the rest frame using radial velocities derived from this study. The upper panel compares the spectra of a single star and a binary star in the saturated regime, with comparable rotation and orbital periods of 0.53 days. The lower panel presents the spectra of three binary stars in the supersaturated, saturated, and unsaturated regimes, with orbital periods of 0.29, 0.53, and 5.95 days, respectively. Among the four unique stars shown (the binary star in the saturated regime appears in both panels), all have similar \teff\ (6227–6374 K) and \logg\ (4.25–4.32) values adopted from the LAMOST LRS Stellar Parameter Catalog. The LAMOST \texttt{obsid} of each spectrum is annotated in the figure. These four stars are marked with green asterisks in Fig.~2.\label{fig:examplespectra}}
\end{figure}

In addition, using the LAMOST \teff, \logg, and [Fe/H] values, we estimated the masses for 74 SB2 systems (Southworth (2015)\cite{southworth2015} and Xiong et al. (2023)\cite{xiong2023}), treating them as single stars, using the \texttt{isoclassify} method. These mass estimates were found to be consistent with the dynamical masses of the primary stars in the SB2 systems, with an offset of 0.03~\msun\ and a standard deviation of 0.20~\msun, corresponding to a scatter of 15.9\% (see Extended Data Fig.~\ref{fig:loggcomp}). Furthermore, this scatter is broadly consistent with the median formal mass uncertainty (20.3\%) reported by \texttt{isoclassify}. Since the mass uncertainties from \texttt{isoclassify} are likely overestimated, we consider the 30\% threshold used in this work reasonable for ensuring a sizable sample. Overall, this result indicates that the mass estimates for binary stars derived using \texttt{isoclassify} are reliable and can be used to approximate the masses of the primaries in \gaia\ binary systems.\\

\noindent\textbf{Validation of LAMOST atmospheric parameters} 
We validate the \teff\ and \logg\ values of unresolved binaries from the LAMOST DR9v2.0 LRS Stellar Parameter Catalog by comparing them to validation values for the primary stars in SB2 binaries, derived using radial velocities (RVs) and light curves. Specifically, we compile 74 SB2 binary systems from Southworth (2015)\cite{southworth2015} and Xiong et al. (2023)\cite{xiong2023}, which also have \teff\ and \logg\ values from LAMOST. The validation \teff\ values\cite{southworth2015,xiong2023} are estimated from spectra observed near the secondary minimum eclipses, minimizing flux contamination from companions. The validation \logg\ values\cite{southworth2015,xiong2023} are derived from model-independent dynamical masses and radii of the primary stars, with uncertainties below 5\%\cite{southworth2015,xiong2023}. In Extended Data Fig.~\ref{fig:loggcomp}, we show that the LAMOST \teff\ and \logg\ values are consistent with the validation values, with an offset of 9~K and a standard deviation of 304~K for \teff, and an offset of 0.00~dex and a standard deviation of 0.12~dex for \logg. This consistency suggests that the LAMOST \teff\ and \logg\ values can be confidently attributed to the primary stars of the SB2 and SB1 binaries and are not substantially affected by light contamination from the companions.

In particular, El-Badry et al. (2018)\cite{el-badry2018} performed experiments with synthetic spectra to investigate systematic biases in atmospheric parameters for unresolved main-sequence binaries analysed with single-star models. They simulated spectra similar to those collected by surveys such as APOGEE, GALAH, and LAMOST. Their results showed that treating an unresolved binary as a single star introduced typical errors of $<$ 100K in \teff, $<$ 0.1dex in \logg, and $<$ 0.05~dex in [\text{Fe/H}] for low-resolution LAMOST spectra. These systematic errors are comparable to measurement uncertainties and align with the validation using SB2 binaries discussed above. Therefore, for all the \gaia\ systems investigated in this work—which are unresolved\cite{arenou2023}—we assume that the values of \teff, \shk, and mass, derived from LAMOST DR9 spectra, correspond to the primary (i.e., brightest) components.\\

\noindent\textbf{Impact of Secondary Components on Activity Measurements}
Among the circularised binaries with orbital periods less than 1 day, where we observe enhanced activity and supersaturation, the fraction of double-lined spectroscopic binaries (SB2) is only 0.17\%. This small proportion indicates that our conclusion on the enhanced activity and supersaturation is not affected by the presence of SB2 systems. Furthermore, among these circularised binaries, 83\% are classified as contact binaries. According to Eggleton (1983)\cite{eggleton1983} and assuming a mass ratio of approximately 0.11\cite{Pesta2023}, the corresponding radius ratio is 2.68 and the luminosity ratio is 7.20, considering a constant temperature across the entire shared surface. This luminosity difference implies that the flux near the Ca II HK lines is dominated by the primary star. Therefore, our activity measurements can be reasonably attributed to the primary stars.

In addition, during the common envelope phase, the effective temperatures of the two components are nearly equal. As a result, the attribution of flux in the Ca II H \& K lines from one component or the other makes little difference. For systems with a mass ratio close to unity, where the Roche lobes are approximately equal, it is important to note that there are effectively two nearly identical stars, not just one, both rotating with the same period as the orbital period. This configuration affects the total observed flux. However, it does not affect \shk, which is normalised to the continuum flux. In the case of two stars with the same \teff, \logrhk would provide an average activity level of the two stars, without introducing effects related to the radiating surface(s). The average activity between the two stars approximately corresponds to that of a single star if the spectra of the two stars are similar, which is expected for the binary stars in our sample that share a common envelope with similar \teff and, especially, comparable masses.

\section*{Data Availability}
The data required to reproduce the figures presented in this work are publicly available from our GitHub repository at \href{https://github.com/Jieyu126/CloseBinaryActivity.git}{this link}. The low and medium resolution spectra, spectral-type classifications, and stellar atmospheric parameters used in this study are available from the LAMOST DR9 survey, accessible \href{http://www.lamost.org/dr9/v2.0/catalogue}{here}. Orbital solutions of binary systems from the \gaia DR3 catalog can be accessed through \href{https://gea.esac.esa.int/archive/}{\gaia\ archive}.

\section*{Code Availability}
The code used in this work is publicly available from our GitHub repository at this \href{https://github.com/Jieyu126/CloseBinaryActivity.git}{link}.

\section*{Acknowledgements}
We thank Jan Henneco, Daniel Huber, Xing Wei, Alexander Shapiro, Zhao Guo, and Tomasz Rozanski for discussions. We also acknowledge Xiang-Song Fang for providing data published in Fang et al. (2018)\cite{fang2018}. We extend our thanks to Orsola De Marco for helping us understand dynamo processes and for her exceptional facilitation of collaboration. J.Y. gratefully acknowledges the tremendous support provided by the entire TOS group at HITS, Heidelberg, during a three-month stay in 2023. J.Y., C.G., R.H.C., and L.G. acknowledge support from ERC Synergy Grant WHOLE SUN 810218. J.Y. and L.G. acknowledge PLATO grants from the German Aerospace Center (DLR 50OO1501) and from the Max Planck Society. We acknowledge funding from the ERC Consolidator Grant DipolarSound (grant agreement \# 101000296). Z.H. acknowledges support from the Natural Science Foundation of China (grant nos 12288102). Y.S.T. acknowledges financial support from the Australian Research Council through DECRA Fellowship DE220101520. J.N. acknowledges support from US National Science Foundation grants AST-2009713 and AST-2319326. S.L.B. and J.Y. acknowledge the Joint Research Fund in Astronomy (U2031203) under a cooperative agreement between the National Natural Science Foundation of China (NSFC) and the Chinese Academy of Sciences (CAS). This work has made use of data from the European Space Agency (ESA) mission \gaia\ (http://www.cosmos.esa.int/gaia), processed by the \gaia\ Data Processing and Analysis Consortium (DPAC; http://www.cosmos.esa.int/web/gaia/dpac/consortium). Funding for the DPAC has been provided by national institutions, in particular, the institutions participating in the \gaia\ Multilateral Agreement. Guoshoujing Telescope (the Large Sky Area Multi-Object Fiber Spectroscopic Telescope, LAMOST) is a National Major Scientific Project built by the Chinese Academy of Sciences. Funding for the project has been provided by the National Development and Reform Commission. LAMOST is operated and managed by the National Astronomical Observatories, Chinese Academy of Sciences.

\section*{Author Contributions}
J.Y. led the project, developed the data reduction package, and conducted the primary analysis, interpretation, and writing. C.G. proposed the S index measurement methods using LAMOST spectra; C.G., Y.S.T., and Y.C. analysed LAMOST spectra. S.H., R.C., and T.R.B. contributed to the design of the work. M.B., S.H., and R.H.C. helped in comprehending enhanced binary activity within the context of saturation for single main-sequence stars. J.Y. and Z.H. identified activity enhancements associated with common envelope evolution. R.H.C., S.J.M., Z.H., and J.N. provided theoretical insights into dynamos. C.G. and P.G. helped understand the activity differences between dwarf and giant binary stars. S.H., R.H.C., and T.R.B. contributed to drafting the work. M.B., S.H., R.H.C., T.R.B., and S.J.M. facilitated collaboration. J.Y., C.G., M.B., S.H., R.H.C., P.G., T.R.B., S.J.M., Z.H., Y.S.T., J.T., Y.C., L.G., J.N., and S.L.B. revised the draft.

\section*{Competing Interests}
The authors declare that they have no competing interests.

\section*{Correspondence}
Correspondence should be addressed to JY (jieyuastro@gmail.com).
\clearpage

\beginsupplement %
\section{Extended Data}

\newpage
\end{document}